\documentclass[superscriptaddress,showpacs,showkeys,twocolumn,aps,pra,10pt]{revtex4-1} 
\usepackage{epsfig}
\usepackage{bm}
\usepackage{amsmath}
\usepackage{subfigure}
\usepackage{graphicx}
\usepackage[pdftex]{color}
\usepackage[colorlinks=true,linkcolor=blue,citecolor=blue]{hyperref}
\usepackage{color}
\usepackage{hyperref}
\hypersetup{
    colorlinks=true,       
    linkcolor=cyan,          
    citecolor=magenta,        
    filecolor=magenta,      
    urlcolor=cyan,           
    runcolor=cyan
}

\usepackage{mathtools}
\DeclarePairedDelimiter{\ceil}{\lceil}{\rceil}

\newcommand{\bea}{\begin{eqnarray}}
\newcommand{\eea}{\end{eqnarray}}
\newcommand{\beq}{\begin{equation}}
\newcommand{\eeq}{\end{equation}}
\newcommand{\bes} {\begin{subequations}}
\newcommand{\ees} {\end{subequations}}

\newcommand{\ua}{\uparrow}
\newcommand{\da}{\downarrow}

\newcommand{\ignore}[1]{}

\begin{document}

\title{Driver Hamiltonians for constrained optimization in quantum annealing}

\author{Itay Hen}
\affiliation{Information Sciences Institute, University of Southern California, Marina del Rey, CA 90292, USA}
\affiliation{Center for Quantum Information Science \& Technology, University of Southern California, Los Angeles, California 90089, USA}
\email{itayhen@isi.edu}
\author{Marcelo S. \surname{Sarandy}}
\email{msarandy@if.uff.br}
\affiliation{Instituto de F\'{\i}sica, Universidade Federal Fluminense, Av. Gal. Milton Tavares de Souza s/n, Gragoat\'a, 
24210-346, Niter\'oi, RJ, Brazil.}
\affiliation{Ming Hsieh Department of Electrical Engineering,
University of Southern California, Los Angeles, California 90089, USA}

\begin{abstract}
One of the current major challenges surrounding the use of quantum annealers for solving practical optimization problems 
is their inability to encode even moderately sized problems---the main reason for this being the rigid layout of their quantum bits 
as well as their sparse connectivity. In particular, the implementation of constraints has become a major bottleneck in the embedding of practical problems, 
because the latter is typically achieved by adding harmful penalty terms to the problem Hamiltonian --- a technique that often requires an `all-to-all' connectivity between the qubits. 
Recently, a novel technique designed to obviate the need for penalty terms was suggested; it is based on the construction of driver Hamiltonians that commute with the constraints of the problem, rendering 
the latter constants of motion. In this work we propose general guidelines for the construction of such driver Hamiltonians given an arbitrary set of constraints. We illustrate the broad applicability of our method by analyzing several diverse examples, namely, 
graph isomorphism, not-all-equal 3SAT, and the so-called Lechner, Hauke and Zoller constraints. We also discuss the significance of our approach in the context of current and  
future experimental quantum annealers. 
\end{abstract}


\maketitle
\section{Introduction}

Quantum annealing (QA)~\cite{kadowaki:98,farhi:01} is a physical approach for optimization that 
utilizes gradually decreasing quantum fluctuations to traverse barriers in the energy landscapes 
of complicated cost functions in search for global minima. As an inherently quantum technique, QA 
is expected to solve combinatorial optimization problems faster than traditional `classical' 
algorithms~\cite{young:08,young:10,hen:11,hen:12,farhi:12}. Recent advances in quantum technology 
have led to the manufacture of the first commercially available programmable quantum annealers 
containing hundreds of quantum bits (qubits)~\cite{johnson:11,berkley:13}. This has prompted a renewed 
interest in schemes for the encoding of real-life problems, and the exciting possibility 
that real quantum devices could solve classically intractable problems of practical importance. 

One of the main advantages of QA is that it offers a very natural approach to solving discrete optimization problems. 
Within the QA framework (often interchangeably referred to as quantum adiabatic optimization), the solution of an
optimization problem is encoded in the ground state of a problem Hamiltonian $H_p$. The encoding is normally readily 
carried out by expressing the problem in terms of an Ising Hamiltonian, which can be interpreted in a simple physical way as 
interacting magnetic dipoles subjected to local magnetic fields. 
To find a minimizing configuration of the problem Hamiltonian, QA prescribes the following course of 
action. As a first step, the system is prepared in the ground state of an initial Hamiltonian $H_d$, commonly referred to 
as the driver Hamiltonian, which must not commute with the problem Hamiltonian $H_p$. The ground state of $H_d$ 
is assumed to be unique and easily preparable. As a next step, the Hamiltonian is 
slowly varied from $H_d$ to $H_p$, normally via the linear interpolation
\bea
\label{eq:hs}
H(s)=s H_p +(1-s) H_d \,,
\eea
where $s(t)$ is the normalized time, with $0 \le s \le 1$ varying smoothly  
between $0$ at $t=0$ to $1$ at time $t=\mathcal{T}$. If the process is performed slowly 
enough, the adiabatic theorem of quantum mechanics~\cite{kato:51,messiah:62} ensures that the system stays close to the ground state of the instantaneous Hamiltonian throughout the evolution, so that one 
finally obtains a state close to the ground state of $H_p$.  At this point, measuring the state will give the 
solution of the original problem with high probability. The running time $\mathcal{T}$ of the algorithm 
determines the efficiency, or complexity, of the algorithm and should be large compared to the inverse 
of a power of the minimum gap~\cite{kato:51,messiah:62, jansen:07,lidarGap}. 

In recent years, it became clear that while QA devices are naturally set up to solve unconstrained 
optimization problems, they are severely limited when it comes to solving problems that involve constraints, 
i.e., when the search space is restricted to a subset of all possible input configurations (normally specified 
by a set of linear equations). The standard approach to imposing these constraints consists of squaring the 
constraint equations and adding them as penalties to the objective cost function with a penalty factor, 
transforming the constrained problem into an unconstrained one~\cite{gaitan:11,gaitan:12,gaitan:14,rieffel:14,lucas:14,rieffel:15}. 
In this approach, the problem Hamiltonian $H_p$ is modified to 
\begin{equation}
H'_p=H_p + \sum_j \alpha_j H^{\text{pen}}_j , 
\end{equation}
 where $H^{\text{pen}}_j$ is defined as 
\begin{equation}
H^{\text{pen}}_j=[C_j(\{\sigma^z_i\})-c_j]^2 \,.
\end{equation}
Here, $C= \{C_j(\{\sigma^z_i\})\}$ denotes a set of constraint operators, and $\{c_j\}$ are constants. The factors $\alpha_j$ are  positive constants suitably chosen to ensure that the ground state of the modified problem Hamiltonian corresponds to that of the original one.
 
However, the addition of penalty terms to the problem Hamiltonian for imposing constraints is very often detrimental to the performance of quantum annealers in 
several ways. First and foremost, it requires many additional interactions to the 
problem Hamiltonian (typically connecting distant neighbors on the hardware lattice). Since actual devices 
cannot accommodate these, costly minor embedding techniques must be employed~\cite{choi:08,vinci:15}. 
Furthermore, the requirement that ground states of $H_p$ map to those of $H'_p$ normally introduces 
 `extra energy scales' to the cost function (encompassed in the values chosen for $\{\alpha_j\}$), which in practice translates to increased error levels in the encoding 
of the couplings (see Ref.~\cite{CQA} for a more detailed discussion). 

To deal with these physical limitations, a novel approach to solving constrained optimization 
problems via quantum annealing, called constrained quantum annealing (CQA), has been recently proposed~\cite{CQA}. 
Within this approach, constraints are enforced via an appropriate choice of driver Hamiltonian, namely, a driver that commutes 
with the constraint operators, i.e.,
\beq \label{eq:hdcj}
[H_d,C_j(\{\sigma^z_i\})]=0 \quad  \forall j \,,
\eeq
but not with the problem Hamiltonian, which turns the constraints into naturally satisfied, or conserved, 
physical quantities. 

Clearly, the tailoring of driver Hamiltonians for a given constraint or set of constraints is far from trivial, and it is unclear at first how easy it is to do so for arbitrary constraints. 
Here, we address, and to some extent resolve, the above matter by
providing general guidelines for constructing appropriate driver Hamiltonians for constrained optimization problems. We further illustrate 
the applicability of these guidelines through several relevant examples, such as graph isomorphism, not-all-equal three-satisfiability (NAE3SAT), 
and the `cycle' constraints introduced in a recent paper on the embedding of fully connected graphs by Lechner, Hauke and Zoller~\cite{LHZ}. 
As we show, the proposed driver Hamiltonians that we construct may in general contain multi-local terms. We therefore also discuss the experimental 
implications of our method and its feasibility in actual near-future quantum annealing devices.

\section{Guidelines to choosing driver Hamiltonians in CQA}

Let us consider a classical problem Hamiltonian $H_p(\{\sigma^z_i\})$ 
and a set of (classical) constraint operators $C= \{C_j(\{\sigma^z_i\})\}$, all of which are functions of the set $\{\sigma^z_i\}$ of Pauli-z operators representing the classical variables of the problem. In what follows, we propose the general principles to construct 
driver Hamiltonians that impose these constraints. 

\subsection{Construction of the driver Hamiltonian}

We begin by observing that, 
since the driver Hamiltonian must not commute 
with classical problem Hamiltonians, it must be composed of off-diagonal operators. 
On the other hand, these terms must commute 
with the constraint operators in order to ensure that the latter are associated with conserved charges~\cite{CQA}. 
To this end, we require that the various terms in the driver Hamiltonian provide a `hopping mechanism' between all the constraint-satisfying (henceforth, allowed) configurations such that the hopping terms 
preserve the desired set $C$ of constraints. Specifically, the driver Hamiltonian must consist of a 
{\it minimal yet complete} set of basic hopping terms, defined as operators that, when acting on one 
allowed configuration, will yield another allowed configuration. 
The hopping terms should therefore be as local as possible, i.e., acting on as few particles as possible---a property that will also render them experimentally more feasible. The second condition on the hopping terms is that they form a complete set in the sense that any allowed configuration is reachable from any other by a sequence of hops. 
Furthermore, hopping terms should never yield forbidden configurations. 

As a final step, the driver 
Hamiltonian is taken to be a linear combination of all terms in the set, with its ground state (in the charge 
sector dictated by the constraint) being a superposition of all allowed configurations. 

\subsection{Setting up the initial ground state\label{sec:ig}}

Setting up the initial state of the system to be the ground state of the driver Hamiltonian 
in the relevant sector $\langle C(\{\sigma^z_i\})\rangle_{t=0}=c$ ensures that the evolution
naturally takes place in the subspace of the allowed configurations of the optimization 
problem, namely, those configurations that automatically obey the constraints of the system.
If however the ground state in the desired charge sector is not the global ground state, 
and the process via which the initial state is prepared (be it the cooling 
down of the system or any other process) does not respect the symmetries of the 
driver Hamiltonian, 
the preparation of the initial state could pose a difficulty, as in this case, the global ground 
state of the system would be favored over the ground state of the desired sector.
This complication may be resolved by adding a 
diagonal `auxiliary correction', $H_{\text{aux}}$, to 
the driver Hamiltonian, modifying the driver Hamiltonian to $H'_d =H_d + H_{\text{aux}}$, where $H_{\text{aux}}$ 
is a linear combination of the constraints 
\beq
H_{\text{aux}} = - \sum_j B_j C_j(\{\sigma^z_i\}), \label{constH}
\eeq
with suitably chosen coefficients $B_j$. Since $H_{\text{aux}}$ is diagonal, it automatically commutes with the problem Hamiltonian while also commuting with $H_d$ [as per Eq.~(\ref{eq:hdcj})]. The inclusion of an auxiliary Hamiltonian with appropriately chosen coefficients $B_j$ can be used to differentially penalize the various charge sectors of $H_d$ such that the relevant charge sector would contain the global ground state~\footnote{In cases where the constraint operators themselves $C_j$ contain $n$-body terms, the locality of $H_{\text{aux}}$ would be the same as that of the usual penalty Hamiltonian. Nonetheless, even in this case, $H_{\text{aux}}$ will generally require a sparser connectivity of the hardware graph.}.
It is important to note however that the inclusion of $H_{\text{aux}}$ of the above form does not always guarantee the existence of values $B_j$ for every charge sector. 

An alternative to the above approach would be to set up the ground state 
using a different diagonal Hamiltonian $H_{\text{aux}}$, whose global ground state is the 
desired one and before the annealing process begins sharply turn off $H_{\text{aux}}$ while turning 
on $H_{d}$.

\subsection{Simple examples}

Before illustrating the application of the above guidelines to nontrivially constrained optimization problems, 
let us consider a couple of special cases --- the first being that of no constraints. Here, the allowed set of 
configurations is the entire set of computational basis states. Thus, the set of basic steps that transform 
one configuration to the next is that composed of all single creation and annihilation spin terms, namely, 
$\{ \sigma^{\pm}_i = (\sigma^x_i \pm i \sigma^y_i) \}$ or rotations thereof. In this case, the 
usual driver Hamiltonian is indeed normally taken to be the Hermitian complete combination of creation 
and annihilation operators provided by the homogeneous transverse-field Hamiltonian 
$H_d= - \sum_i \sigma^{x}_i$. 

A somewhat less trivial example which adheres to the above principles has been introduced and 
discussed in Ref.~\cite{CQA}, where the the constraint $\langle \sum_{i=1}^n \sigma_i^z \rangle = c$ has been 
studied in the context of the graph partitioning problem. There, the suggested driver Hamiltonian was
 \bea
 \label{eq:driver}
 H_d&=&- J \sum_{i=1}^n \left( \sigma_i^x \sigma_{i+1}^x + \sigma_i^y \sigma_{i+1}^y\right) \\\nonumber
 &=& -\frac{J}{2}\sum_{i=1}^n \left(\sigma^+_i \sigma^-_{i+1} + \sigma^-_i \sigma^+_{i+1}\right) \,,
 \eea 
where $J$ sets the energy scale and periodic boundary conditions are adopted, namely,  $\sigma_{n+1}^{x/y}=\sigma_{1}^{x/y}$.
The above driver is a special case of the well-known XY-model~\cite{lieb:61,pasquale:2008}
and can be viewed as a sum of terms that describe the hopping of a particle from an occupied 
site to an unoccupied one. In spin terminology, the driver Hamiltonian consists of moves of the 
form $|\ua \da\rangle \leftrightarrow |\da \ua \rangle$. One can easily observe that this driver also 
naturally follows the above principles, offering hopping terms between allowed configurations, 
conserving `number of particles' or equivalently, the total $z$-magnetization. 

\begin{figure}[ht]
\begin{center}
\includegraphics[width=0.99\columnwidth]{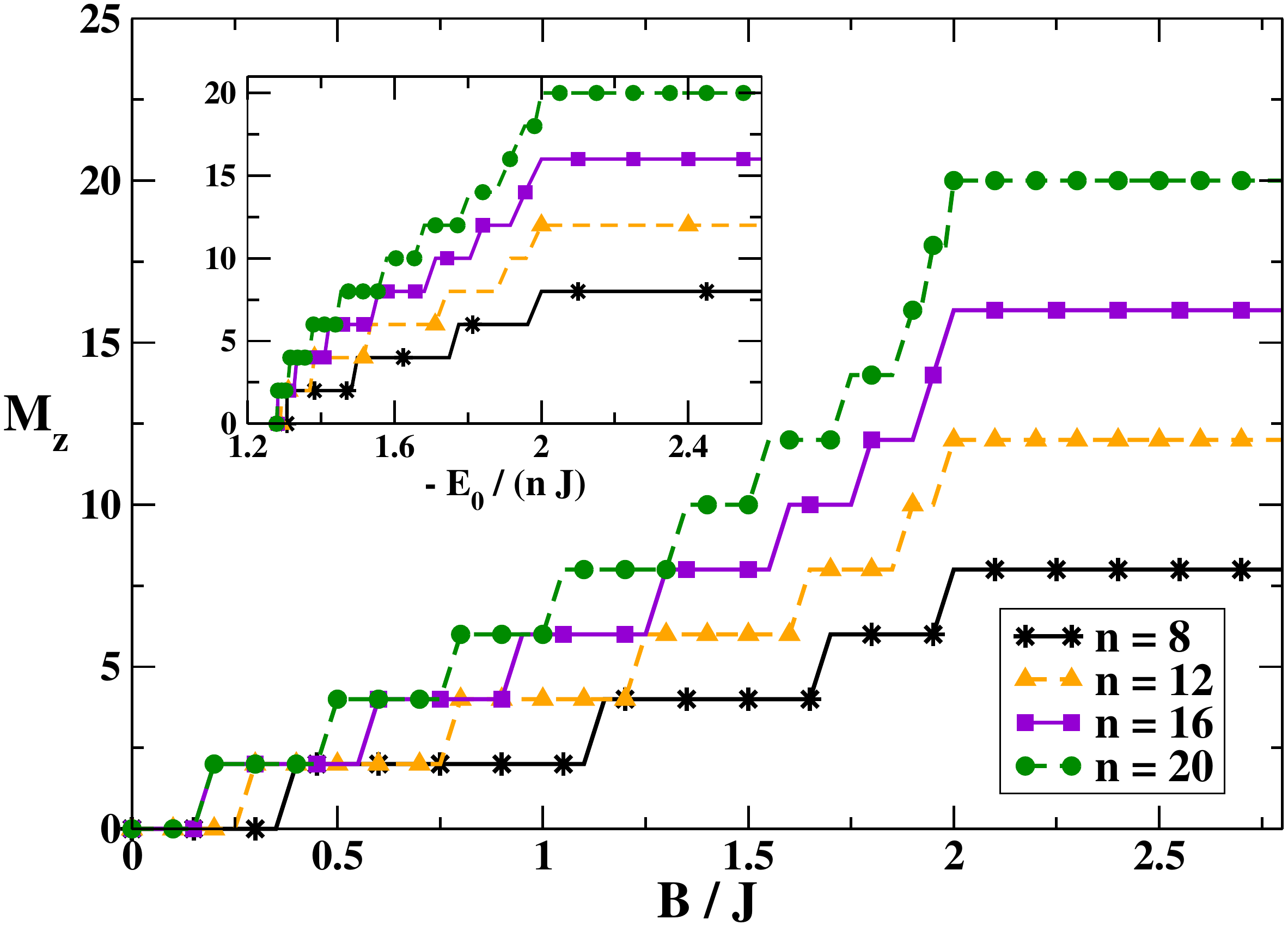}
\caption{(Color online) Magnetization $M_z=\langle \sum_i \sigma_i^z \rangle$ of the global ground state of the XY driver as a function of the 
ratio $B/J$ for periodic chains of various sizes. Ground states with negative magnetization $M_z<0$ are attained similarly for negative $B$ values. {\it Inset}: Magnetization as 
a function of the ratio $-E_0/(nJ)$, where $E_0/n$ is the ground state energy density. }
\label{fig:0}
\end{center}
\end{figure}

The global ground state of the XY Hamiltonian is in the zero-magnetization sector. 
If however the magnetization sector required by the constraint is
$\langle \sum_{i=1}^n \sigma_i^z \rangle = c$ with $c \neq 0$ (an example for that appears in the next section), one can add to the XY driver an auxiliary Hamiltonian \hbox{$H_{\text{aux}}=-B\sum_i\sigma^z_i$}  with a properly chosen coefficient $B$ which would shift the global ground state to the desired magnetization sector. Figure~\ref{fig:0} depicts the relation between the magnetization of the ground state of the XY chain as a function of $B/J$. 
In the inset, we show a similar behavior in terms of the ground state energy density 
$E_0/(J\,n)$.

We now proceed to demonstrate the generality of the suggested guidelines by illustrating their 
applicability in more general constrained optimization problems. 

\section{The graph isomorphism problem}

One of the most notable examples of what may be referred to in the context of quantum annealers as 
a constrained optimization problem is that of graph isomorphism (GI). 
The problem is stated as follows. Given two 
graphs $G_1$ and $G_2$, one must determine whether or not they are isomorphic to
each other, i.e., whether one can be transformed into the other by a relabeling of the vertices.
Very recently, a classical algorithm running in quasi-polynomial time $\exp\left[\log\left(n\right)^{O(1)}\right]$, 
with $n$ denoting the number of vertices, has been proposed for the GI problem by Babai~\cite{Babai:15},  which constitutes a breakthrough in complexity theory. Whether or not further speedup is achievable 
by quantum optimization remains an open question.  
Several techniques have been 
introduced in the recent past to encode GI problems on Ising-type quantum annealers~\cite{hen:12b,gaitan:14,lucas:14}. 
In the canonical approach worked out explicitly by Lucas~\cite{lucas:14}, the mapping of an $n$-vertex graph to an Ising lattice
requires an $n$ by $n$ square grid of spin-1/2 particles, whose corresponding Pauli-z operators are denoted by the doubly-indexed 
$\sigma^z_{i,j}$ with $i,j=1\ldots n$. The computational states $|\uparrow\rangle$ or $|\downarrow\rangle$ at site $(i,j)$ 
indicate whether or not the $i$-th vertex of one graph is to be identified with the $j$-th vertex of the 
second graph. A problem Hamiltonian can then be written in terms of positive energy contributions to bad mappings, i.e., each time an 
edge appears in one graph but not in the other. This yields
\bea\label{eq:hpGILucas}
H_p &=&  \sum_{ij\notin E_1} \sum_{i'j'\in E_2} \frac{(1+\sigma^z_{i,i'})}{2}\frac{(1+\sigma^z_{j,j'})}{2}  
\nonumber \\
&+& \sum_{ij\in E_1} \sum_{i'j'\notin E_2} \frac{(1+\sigma^z_{i,i'})}{2}\frac{(1+\sigma^z_{j,j'})}{2}  \,,
\eea
where $E_1$ and $E_2$ are the edge sets of $G_1$ and $G_2$, respectively. 
Additionally, $2 n$ constraints ensuring that the mapping is {\it bijective} are required, namely,   
$C^{(1)}_j=\sum_i \left( 1+\sigma^z_{i,j} \right)/2 =1$ and $C^{(2)}_j=\sum_i \left(1+ \sigma^z_{j,i}\right)/2=1$, for each $j \in \{1,\ldots,n\}$.
The solutions of the GI problem are then contained in the ground state $|\psi\rangle$ of $H_p$ under the constraints 
$C=\{C^{(1)}_j,C^{(2)}_j\}$ obeying $H_p |\psi\rangle = 0$. 

In the standard approach, the $2n$ constraints are translated 
into penalty terms in the problem Hamiltonian, explicitly,
\beq \label{HpenaltyGI}
H^{\text{pen}}=\sum_i \Big[ \big( \sum_{j} \frac{\left( 1+\sigma^z_{i,j} \right)}{2} - 1 \big)^2 + \big( \sum_{j} \frac{\left(1+ \sigma^z_{j,i}\right)}{2} - 1\big)^2 \Big] \,.
\eeq

\begin{figure}[ht]
\begin{center}
\includegraphics[width=0.99\columnwidth]{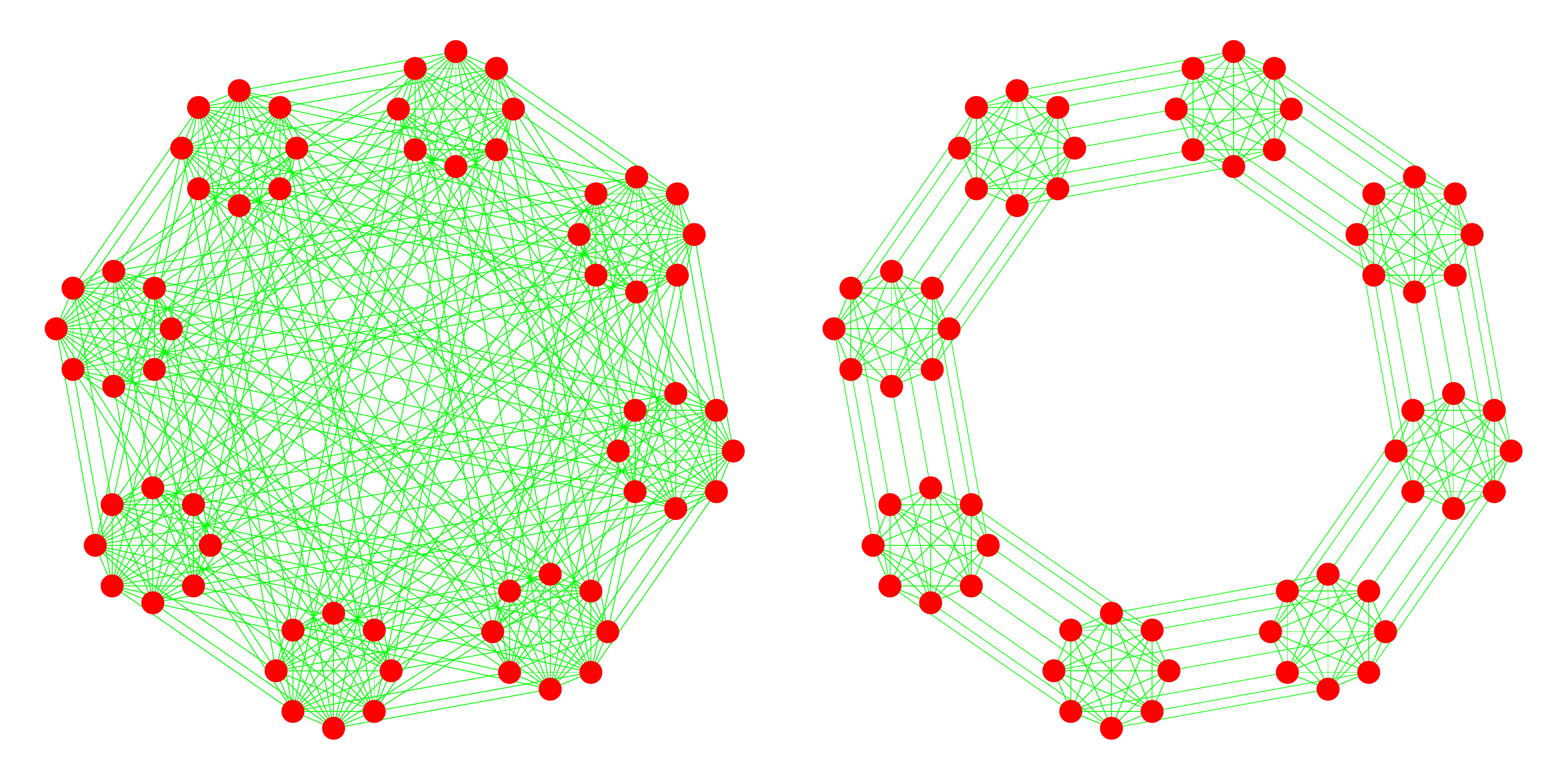}
\caption{(Color online) Required connectivities in the GI problem: an 8-vertex example. {\it Left}: 
Connectivities in the standard approach. {\it Right}: Partial removal of connectivities through the use of driver XY Hamiltonians. 
}
\label{fig:GIconn}
\end{center}
\end{figure}

As noted earlier, the addition of penalties can be detrimental to the embeddability of GI problems on experimental quantum annealers. This was, in fact, illustrated in a recent experiment which examined the embedding of GI instances on the so-called D-Wave Two experimental quantum annealing processor, where the resources required for the encoding of instances on the sparsely connected 504-qubit chip allow for the embedding of graphs of at most 6 vertices~\cite{zick:15}. The reason for this highly inefficient embedding stems partly from the penalty terms given in Eq.~(\ref{HpenaltyGI})---each of the $2n$ constraints requires the existence of an $n$-qubit clique, i.e., $n(n-1)/2$ edges; the total number of required edges thus amounts to $N_E = n^2(n-1)$. This is illustrated in Fig.~\ref{fig:GIconn} (left). 

In the next subsections, we will describe several 
alternatives to the above penalty-based embedding that utilize the freedom of choosing suitable driver Hamiltonians in order to substantially reduce the amount of resources required for the encoding of GI instances.

\subsection{Partial removal of penalty terms}

A partial removal of penalty terms can be achieved by the simple observation that each of the two sets of 
constraints $\{ C_j^{(1)}\}$ and $\{ C_j^{(2)}\}$ contain disjoint sets of bits. Since the constraints are of the `total $z$-magnetization' 
type, one can remove either set by setting up a sequence of cyclic XY 
driver Hamiltonians, of the type introduced in Eq.~(\ref{eq:driver}). An XY driver can be used for each of the $n$ constraints in, say, set $\{C_j^{(1)}\}$, forming $n$ independent $n$-qubit cycles, each spanning one row on the square grid. 
Specifically, consider the driver Hamiltonian
 \beq
 H_d=- \sum_{i=1}^n \sum_{j=1}^{n} \left( \sigma_{i,j}^x \sigma_{i,j+1}^x + \sigma_{i,j}^y \sigma_{i,j+1}^y\right) \,,
\eeq
augmented with the periodic boundary $\sigma_{i,n+1}^{x/y}=\sigma_{i,1}^{x/y}$ and  
$\sigma_{n+1,j}^{x/y}=\sigma_{1,j}^{x/y}$. Since the total magnetization in each row is $M_z=n -2$, 
 the search space here is one in which all the spins belonging to the same vertex but one, must 
point down (imposing one target vertex in $G_2$ per each vertex in $G_1$). The above driver thus preserves the 
magnetization in each row on the square grid. This immediately reduces the amount of 
required connections from $N_E = n^2(n-1)$ to $N_E^\prime=n^2(n-1)/2+n^2$, as illustrated in 
Fig.~\ref{fig:GIconn} (right). In this case, the second set of constraints will be imposed as before in the form of
penalty terms, and the evolution of the system will take place in the subspace spanned by the allowed states obeying all first 
$n$ constraints. As was discussed in the previous section, an auxiliary Hamiltonian \hbox{$H_{\text{aux}}=-B \sum_i \sigma^z_i$} can be added to the driver in order to make the lowest-lying state of the $M_z=n-2$ sector the global ground state. 

\subsection{Four-body terms}

\begin{figure}[ht]
\begin{center}
\includegraphics[scale=0.45]{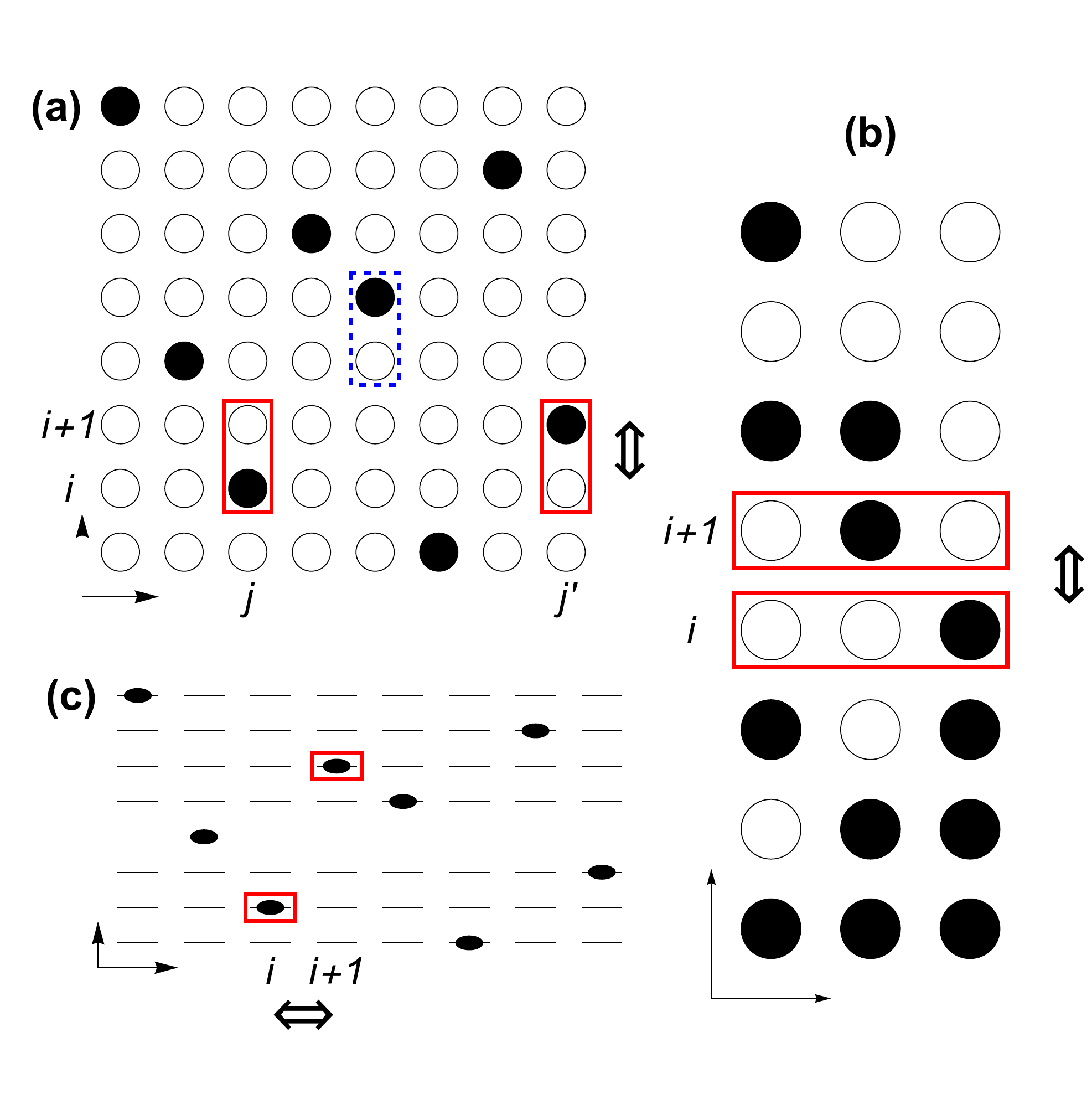}
\caption{(Color online) Different hopping term types in the graph isomorphism driver: An 8-vertex example. The solid (red) 
rectangles denote the particles involved in a single hopping term. The dashed (blue) rectangle denotes a hopping term in the 
case of partial removal of the constraints. {\it (a)} An empty circle denotes a down spin (or a vacant site), whereas a full 
circle denotes an up spin (or an occupied site). A hopping term here consists of four-body terms. {\it (b)} An $n$ by 
$\ceil{\log_2 n}$ grid with local hopping terms swapping neighboring rows. {\it (c)} An $n$-qudit setup with $n$ levels 
each. The hopping terms are two-body and local, swapping the modes of neighboring particles. }
\label{fig:GIhopp}
\end{center}
\end{figure}

Let us now employ our hopping mechanism approach to the GI problem in order to remove all penalty terms.
Here, as we shall see, the removal of all penalties comes at 
the cost of introducing four-body terms in the driver Hamiltonian. In this scenario, the minimal set of hopping terms required to hop from one allowed configuration to another is given by terms of the type:
\bea \label{Hd4b}
H_d=-\sum_{i,j}\sum_{j<j'} \Big( &|\ua\rangle&\langle \da|_{(i,j)} \otimes  |\da\rangle\langle \ua|_{(i+1,j)} \\\nonumber
\otimes &|\ua\rangle&\langle \da|_{(i,j')} \otimes  |\da\rangle\langle \ua|_{(i+1,j')}
+\text{c.c.} \Big) \,,
\eea
where $\text{c.c.}$ denotes complex conjugate terms. 
The interpretation of the Hamiltonian above in terms of hopping 
particles is illustrated in Fig.~\ref{fig:GIhopp}(a) where an 8-vertex graph is considered. The hopping terms swap the location of particles (or up spins) in neighboring rows, an operation that preserves the total $z$-magnetization $M_z=\sum_i\sigma^z_i$ in each row and column on the square grid. 
We note here that similar moves have also been considered in Ref.~\cite{martonak:04} in the context of the traveling salesman problem.
By rewriting $H_d$ in Eq.~(\ref{Hd4b}) in terms of Pauli operators, we get
\beq
H_d=-\sum_{i,j}\sum_{j<j'} \left( \sigma^+_{(i,j)}\sigma^-_{(i+1,j)}\sigma^+_{(i+1,j')}\sigma^-_{(i,j')} +c.c. \right) .
\eeq
Note that $H_d$ above involves four-body terms. This is because commutation with the GI constraints requires conservation of 
magnetization along the $z$ direction in the individual rows and columns on the square grid of spins. In particular, four-body interactions are enough to implement the solution of 
any $n$-vertex GI problem; i.e., the non-locality of the interaction does not scale with the size of the problem. 
The advantage of the current approach is that, at the cost of a four-body quantum driver Hamiltonian, no additional constraints 
are needed in the problem Hamiltonian, meaning that $N_E=O(n^2)$ edges are already enough to embed an $n$-vertex graph. 

\subsection{Resource-efficient many-body hops}
We now proceed to introduce other setups which allow for the removal of all penalty terms from the problem Hamiltonian. The setups we propose here are more efficient in terms of physical resources, 
requiring only $n\ceil{\log_2 n}$ physical qubits. Here, the mapping between any vertices of $G_1$ and $G_2$ is binary encoded in $\ceil{\log_2 n}$ qubits (this representation 
shares some similarities with the encoding introduced in Ref.~\cite{gaitan:14}). This encoding is illustrated in Fig.~\ref{fig:GIhopp}(b).
In this case, the problem Hamiltonian is given by
\bea\label{eq:hpGIqudit}
H_p &=&  \sum_{ij\notin E_1} \sum_{i'j'\in E_2} |i^\prime\rangle \langle i^\prime |_i \otimes |j^\prime \rangle \langle j^\prime |_j  
\nonumber \\
&+& \sum_{ij\notin E_1} \sum_{i'j'\in E_2} |i\rangle \langle i |_{i^\prime} \otimes |j\rangle \langle j |_{j^\prime}    \,,
\eea
where the $\ceil{\log_2 n}$-qubit-long binary representations of the $G_2$ nodes are denoted here by $|i\rangle$ with $i=\{0\ldots n-1\}$. 
The above terms penalize edges that exist in one graph but not in the other.  
With the above encoding, a suitable driver Hamiltonian is given by
\beq\label{eq:hdGIqudit}
H_d=-\sum_{i=0}^{n-1} \sum_{j=0}^{n-1} \sum_{j' \neq j} |j\rangle \langle j' |_i \otimes |j'\rangle \langle j |_{i+1} \,,
\eeq
where the various terms swap neighboring blocks of $\ceil{\log_2 n}$ qubits. Note that the reduction in the number of physical qubits here comes at the price 
of the existence of many-body terms that scale logarithmically with the size of the input graphs. This is because the hops depict the swapping of neighboring $\ceil{\log_2 n}$-qubit blocks [see Fig.~\ref{fig:GIconn}(b)].

\subsection{Two-body local terms using $n$-level qudits}

Last, we discuss an implementation of the hopping technique which removes the penalty terms altogether while also requiring only two-body local interactions in the driver Hamiltonian. This encoding makes use of the same driver and problem Hamiltonians as in the `many body' case above, namely Eq.~(\ref{eq:hpGIqudit}) and Eq.~(\ref{eq:hdGIqudit}), respectively. Now, however, the states $|j\rangle$ no longer represent a logarithmic number of qubits but rather an $n$-level qudit. This scenario is illustrated in Fig.~\ref{fig:GIhopp}(c).
This Hamiltonian may be implemented in, e.g., the framework of linear optics quantum computing (LOQC) where the qudit levels correspond to second-quantized modes of photons. Here, the driver Hamiltonian contains level- or mode-swapping operators between neighboring qudits, making the driver both local and two-body. 

We conclude this section by noting the existence of numerous NP-hard or NP-complete problems that may readily utilize the encoding methods discussed above in the context of the GI problem. Among these are the traveling salesman problem and problems in planning and scheduling, which are also based on finding optimal permutations of bijective maps.

\section{Not-all-equal 3SAT}

The not-all-equal $3$SAT (NAE3SAT) problem is a special type of a bigger class of constraint satisfaction problems in which one has to determine the existence of satisfying $n$-bit assignments given a list of $m$ logical conditions, or clauses, each defined on a small number of randomly chosen bits. This problem and variants thereof have recently become a focus of much interest in the context of experimental quantum annealers~\cite{nae3satKing,hen:15}.

In NAE3SAT, each clause consists of three bits, and a clause is satisfied for six of the $2^3=8$ possible configurations, with a remaining
pair $\{j,\bar{j}\}$ of violating configurations, where $j$ denotes one 3-bit configuration and $\bar{j}$ its global negation.
A configuration of the bits (spins) is a satisfying assignment if it satisfies all the clauses.
In the standard encoding of this type of problem in the context of quantum annealing, each bit variable
is represented in the Hamiltonian by the $z$-component of a Pauli
matrix, $\sigma_i^z$, where $i$ labels the spin.  Each clause is thus
converted to an energy function which depends on the spins associated with the
clause, such that the energy is zero if the clause is satisfied and is
positive if it is not.  The problem Hamiltonian then becomes $H_p = \sum_{m=1}^{M} H^{(m)}$, 
where $m$ is the clause index and $H^{(m)}$ is the energy associated with
the clause and involves only the spins belonging to it. 
The NAE3SAT clause Hamiltonian $H^{(m)}$ can succinctly be written as
$H^{(m)}=|j_m\rangle \langle j_m| + |\bar{j}_m\rangle \langle \bar{j}_m|$, where $|j_m\rangle$ and $|\bar{j}_m\rangle$ are the 3-bit computational basis states corresponding to the two violating configurations of the $m$-th clause. Here, the energy
is zero if the clause is satisfied and is non-vanishing otherwise.

To demonstrate the applicability of our approach, in what follows we shall treat some of the terms in $H_p$ as constraints that are to be converted to conserved 
quantities. Let us denote this set of constraints by $\mathcal{C}$, while the other clauses shall remain part of the Hamiltonian. We shall require that 
the  constraint clauses are mutually disjoint, i.e., operating on disjoint sets of spins. In this case, the problem Hamiltonian would consist 
only of non-constraint clauses---explicitly,  $H_p=\sum_{m \notin  \mathcal{C}} H^{(m)}$---as the other constraints would be naturally
conserved provided a suitable driver is found.
As for the driver Hamiltonian, one may define a single-clause hopping term as the sum of equal probability transitions between the allowed configurations of the clause, explicitly:
\beq
H_d^{(m)}= -\sum_{i \neq j_m,\bar{j}_m} |i\rangle \sum_{i' \neq i,j_m,\bar{j}_m} \langle i'|\,,
\eeq
where $H_d^{(m)}$ acts on the three bits in the clause $m \in  \mathcal{C}$. 
Note that the above driver term requires $3$-local terms such as the problem Hamiltonian clauses $H^{(m)}$.  
For all bits $k$ that are not present in the clauses chosen to be conserved, if there are such, a transverse-field driver will be chosen. 
Therefore, the total driver 
Hamiltonian will be
\begin{equation}
H_d=\sum_{m \in  \mathcal{C}} H_d^{(m)} - \sum_k \sigma^x_k,
\end{equation}
where $k$ labels all the spins {\it not} present in any of the constraint clauses of $\mathcal{C}$.  
The ground state for this driver is simply
\beq
|\psi\rangle = \bigotimes_{m \in \mathcal{C}} \left( \frac1{\sqrt{6}}\sum_{i \neq j_m,\bar{j}_m} |i\rangle \right)  \bigotimes_{k} |+\rangle_k \,.
\eeq

\section{The Lechner, Hauke and Zoller cycles}

Recently, Lechner, Hauke and Zoller (LHZ)~\cite{LHZ} have proposed a quantum annealing architecture in which a classical $n$-bit spin glass with all-to-all connectivity is mapped to a spin glass with $M=n(n-1)/2$ bits and geometrically local interactions. In the LHZ scheme, the bits correspond to products $\sigma^z_i \sigma^z_j$ of the original fully connected problem, and so Ising problem Hamiltonians are mapped onto Hamiltonians of the form $H_p=\sum_{k=1}^{M} J_k \sigma^z_k$. Since these new spin variables are dependent on each other, the new Hamiltonian is subject to $L=M-n$ constraints of the form  
\beq\label{eq:lhzConst}
C_l=\bigotimes_m\sigma^z_{l_m}=1 \,,
\eeq
where $l=1\ldots L$ labels the constraints, and the spins of each constraint (denoted by $l_m$ above) trace overlapping cycles on the LHZ hardware graph~\cite{LHZ}. 

Imposing the above constraints using $L$ penalty Hamiltonians, e.g., by adding $\sim n^2$ penalty terms of the form 
$\alpha_l H^{\text{pen}}_l=\alpha_l (1-C_l)^2$ [or equivalently, $\alpha_l H^{\text{pen}}_l=\alpha_l(1-C_l)$ since $C_l^2=1$] to the problem Hamiltonian, drives up the energy scale of the problem depending on the choice of $\alpha_l$. In practice, this large number of XORSAT-type constraints~\cite{jorg:10} is expected to be detrimental to the performance of any device implementing the above structure. This can be attributed to the intricate energy landscapes generated by the XORSAT terms, which has long been recognized to stymie heuristic optimization algorithms, quantum as well as classical~\cite{guidetti:11,hen:11,farhi:12}. 

To partially remove the LHZ constraints, one may choose, as in the NAE3SAT case discussed in the previous section, a set of non-overlapping cycles $\mathcal{C}$ that will be identified as constraints that are to be removed. As a next step, one would set up a suitable driver Hamiltonian to enable the elimination of these from the problem Hamiltonian. 
A driver Hamiltonian term for a constraint of the form Eq.~(\ref{eq:lhzConst}) above is easily constructed if one notices that the flipping of any two spins in the constraint provides a general hopping mechanism from one allowed configuration to another. This allows for driver Hamiltonian terms of the form
\beq
H_d^{(l)}= - \sum_{m} \sigma^x_{l_m} \sigma^x_{l_{m+1}}\,,
\eeq
for each of the constraints in the set $\mathcal{C}$. This Hamiltonian provides the two bit-flip hopping mechanism between configurations. It has two ground states only one of which corresponds to the correct charge sector $C_l=1$ (the other corresponds to $C_l=-1$). The ground state in the $C_l=1$ sector is the equal superposition of all states that are an even number of spin flips away from the state of all spins pointing  in the positive $z$ direction. 
As was discussed in Sec.~\ref{sec:ig}, applying a small magnetic field $H_{\text{aux}}$ in the $z$-direction easily breaks the degeneracy in favor of the former (correct) configuration [see Eq.(\ref{eq:lhzConst})]. 

The total driver Hamiltonian is therefore
\begin{equation}
H_d=\sum_{l \in  \mathcal{C}} H_d^{(l)} - \sum_k \sigma^x_k,
\end{equation}
where $k$ labels all the spins {\it not} present in any of the constraint clauses of $\mathcal{C}$.  
The above setup allows for only a partial removal of the $L$ constraints as it allows for the elimination of only  non-overlapping cycles. 

Despite the intricate structure of the above system of constraints, one finds that there is indeed a driver Hamiltonian that naturally constricts the evolution of the system to the subspace of allowed configurations and which obviates the need for harmful penalties. To construct it, we first observe that each constraint of the form Eq.~(\ref{eq:lhzConst}) depicts, in fact, a linear equation. Since the constraints are classical, the equation for the $l$-th constraint can be written in the  form $\prod_m s_{l_m}=1$ where $s_{l_m}=\pm 1$ denote binary variables, or Ising spins. Alternatively, the constraints may take the form $\prod_m (-1)^{b_{l_m}} =1$, or equivalently
\beq
\sum_m b_{l_m} = 0 \mod 2 \,,
\eeq
in terms of the Boolean variables $b_{l_m} \in \{0,1\}$. The set of all $L$ constraints therefore represents a linear system of mod-$2$ equations which can be readily solved using, e.g., Gaussian elimination.
Solving for $L$ of the $M$ bits, the obtained solutions are of the form
\beq \label{eq:sl}
s_l = \prod_{m} s_{l_m} \,,
\eeq
where the variables $\{ s_l\}_{l=1}^{L}$ on the left-hand side are the dependent solved-for variables, and the $s_{l_m}$ variables on the right-hand side belong to the set of the $n$ remaining independent variables. 
The above solution suggests that the LHZ constraints may be rewritten accordingly as 
\beq\label{eq:lhzConst2}
C_l=\sigma^z_l \bigotimes_m\sigma^z_{l_m}=1 \,,
\eeq
where $\sigma^z_l$ with $l=1\ldots L$ are the `dependent' operators (one in each constraint) and 
$\{ \sigma^z_{l_m}\}$ are independent operators. 

With the constraints now cast in the above form, the removal of penalty terms can be carried out in one of two ways. One is by back-substituting the solved-for variables into the LHZ problem Hamiltonian \hbox{$H_p=\sum_k J_k \sigma^z_k$}. 
Another approach would be to consider driver Hamiltonian terms composed of products of the form 
\beq\label{eq:hdl}
H_d^{(p)}=\bigotimes_{i \in \mathcal{S}_p} \sigma^x_{i} \bigotimes_{d \in \bar{\mathcal{S}}_p} \sigma^x_{d}\,,
\eeq
where the operators in the first tensor product correspond to independent variables 
and those in the second product associated with dependent variables.  
In order to construct a driver term $H_d^{(p)}$ that commutes with all $L$ constraints in Eq.(\ref{eq:lhzConst2}), the following strategy can be adopted. For any nonempty subset of independent operators $\mathcal{S}_p$, 
the subset of dependent operators $\bar{\mathcal{S}}_p$ can be determined after observing that,   
if the product of the operators in $\mathcal{S}_p$ does not commute with a 
given constraint, the addition of the dependent variable of that constraint to 
$\bar{\mathcal{S}}_p$ rectifies the situation. Conversely, if the product of the operators in $\mathcal{S}_p$ does commute with that given constraint, the dependent variable is not added. 
This scheme ensures that the effective flipping of independent variables necessarily implies the 
flipping of the appropriate dependent variables in such a way that all 
constraints are satisfied. 
We note that since the number of 
dependent variables is $L \sim n^2$, the driver terms may in principle contain highly nonlocal 
contributions. However, this non-locality will depend on the choice of $\mathcal{S}_p$ and may in principle be minimized.   

A driver Hamiltonian consisting of an appropriate linear combination of driver terms of the above form would immediately remove the need for constraint penalties in the problem Hamiltonian. 
Since the eigenspectrum of each driver term $H_d^{(p)}$ splits the Hilbert space into two 
equally sized subspaces, a suitable choice of $M$ linearly independent driver terms will result in a driver with a unique ground state, as required. 

\section{Conclusions}

In this work, we addressed the question of how to tailor driver Hamiltonians to quantum annealing 
processes that aim to solve constrained optimization problems without resorting to the use of penalty terms.
We provided general guidelines for choosing suitable driver Hamiltonians given a constraint or a set of 
constraints and  demonstrated the broad applicability and benefits of constrained quantum annealing (CQA). 

As we have seen, the advantages of using driver Hamiltonians that provide a hopping mechanism between allowed configurations are often considerably more beneficial than existing methods in terms of the resources required for embedding certain problems and obviating the need for harmful penalty terms. However, the locality and hence the experimental feasibility of some of the driver Hamiltonians requires further attention. While for some problems, a specifically tailored driver Hamiltonian is highly nonlocal, in other examples, the necessary off-diagonal terms are not only resource-efficient but also experimentally feasible.
 
It would therefore be of interest to be able to classify constraints according to the locality of the driver Hamiltonian required to conserve them. Specifically, obtaining a class of constraints that can be dealt with using only experimentally feasible two-body geometrically local interactions would be of special practical importance. 
To date, the embedding of practical optimization problems on experimental quantum annealers has been considered impractical, suffering from the detrimental resource requirements of embedding techniques that have so far been required for the imposing of constraints. We hope that this work will motivate experimental engineering of suitably constructed  quantum annealers, eventually bringing closer the feasibility of encoding real-life optimization problems approaching the practical regime. 
 
 \section*{Acknowledgements} 
M. S. S. thanks Daniel Lidar for his 
hospitality at the University of Southern California. 
M. S. S. acknowledges support from the Brazilian agencies CNPq, CAPES, FAPERJ, and the Brazilian National Institute for Science and Technology of Quantum Information (INCT-IQ).

\bibliography{refs}

\end{document}